\documentclass[twocolumn,showpacs,preprintnumbers,amsmath,amssymb,floatfix]{revtex4-1}

\usepackage{epsfig,psfrag}
\usepackage{dcolumn}
\usepackage{bm}
\usepackage{graphicx}
\usepackage{color}

\begin{document}

\title{Electron-phonon scattering in topological insulators}
\author{S{\'e}bastien Giraud and Reinhold Egger}
\affiliation{Institut f\"ur Theoretische Physik, 
Heinrich-Heine-Universit\"at, D-40225  D\"usseldorf, Germany}
\date{\today}

\begin{abstract}
We formulate and apply a theory of electron-phonon interactions 
for the surface state of a strong topological insulator. 
Phonons are modelled using an isotropic elastic continuum theory
with stress-free boundary conditions and interact with the 
Dirac surface fermions via the deformation potential. We
discuss the temperature dependence of the quasi-particle lifetime in
photoemission and of the surface resistivity.
\end{abstract}

\maketitle

\textit{Introduction.---}
One of the presently most active areas in physics is concerned with 
strong topological insulator (TI) materials \cite{hasan,qizhan}.   
In these systems strong spin-orbit couplings cause band inversion 
and a nontrivial topology of the map from momentum to Hilbert space 
\cite{fukane,qiz}. In a TI, as long as time-reversal invariance remains
unbroken, an odd number of massless surface Dirac fermion modes  
is guaranteed despite of the presence of a bulk gap $\Delta_b$.  
The existence of metallic two-dimensional (2D) Dirac surface states 
has been convincingly established using angle-resolved 
photoemission spectroscopy (ARPES) in bismuth selenides \cite{hasan,arpes}.  
Typical reference materials are Bi$_2$Te$_3$ or  Bi$_2$Se$_3$, where 
$\Delta_b\simeq 0.3$~eV allows to observe these phenomena even
at room temperature, but other material classes
have also been predicted to possess a TI phase \cite{hasan}. 
Attempts to observe electronic transport signatures of the 
surface state were only partially successful \cite{butch,ong,analytis} 
since surface effects are easily masked by defect-induced 
residual bulk charge carriers. Anticipating progress in 
achieving better purity, it is important to understand 
what intrinsically limits the surface conductivity and
the integrity of  surface quasi-particles. 
Noting that the large and anisotropic static 
dielectric constant ($\epsilon\approx 50$ to $200$ 
\cite{richter}) implies a drastic reduction of direct 
Coulomb forces or charged impurity potentials, we 
here analyze consequences of the \textit{electron-phonon coupling} 
on the TI surface state.

We focus on long-wavelength acoustic phonons which 
dominate the physics at low energy scales. Previous work on the
thermoelectric properties of Bi$_2$Te$_3$
has demonstrated that despite of the quintuple-layer crystal structure,  
bulk acoustic phonons are reasonably well described as
isotropic elastic continuum \cite{jenkins,huang}, where the
two Lam{\'e} parameters of the theory determine the longitudinal and 
transverse sound velocities, $c_l\simeq 2800$~m/s and 
$c_t\simeq 1600$~m/s, respectively.  
Low-temperature electronic transport is then limited by the 
deformation potential coupling to acoustic phonons, since piezoelectric
couplings are suppressed by inversion symmetry \cite{huang}.
TI experiments have so far only addressed 
the coupling to optical phonons \cite{richter,shahil}, cf.~also studies for Bi 
surfaces \cite{hofmann}. However, massless 2D Dirac fermions are realized
in graphene monolayers as well, where
both theory \cite{dassarma,felix} and experiment \cite{kim} have reported 
consistent results for the temperature ($T$) dependence of the resistivity
($\rho$):  for $T\ll T_{\rm BG}$ (with the Bloch-Gr\"uneisen 
temperature $T_{\rm BG}\equiv 2\hbar k_F c_s/k_B$, Fermi momentum $k_F$, and
sound velocity $c_s$), a $\rho \sim T^4$ scaling is found, 
while $\rho\sim T$ for $T\gg T_{\rm BG}$.
Note that for a 2D electron gas with parabolic dispersion and dominant
deformation potential coupling,
one expects $\rho\sim T^7$ for $T\ll T_{\rm BG}$ \cite{knab}.
Below, we shall discuss the $\rho(T)$ dependence of the TI surface state 
due to the coupling to acoustic phonons in detail.
In addition, phonons are expected to cause quasi-particle decay. This implies,
e.g., a finite ARPES linewidth \cite{eiguren,echenique}, where an 
anomalous behavior was observed in Bi$_2$Se$_3$ \cite{park}.

In this paper, we formulate and study an analytically tractable
effective low-energy theory for the TI surface state 
coupled to acoustic phonons.  The phonon modes are obtained from  
isotropic elastic continuum theory in a half-space 
with stress-free boundary conditions \cite{landau,sirenko}. Their coupling
to the surface fermions is predominantly via the deformation potential 
\cite{thalmeier}.  For concrete numbers, we use published 
\cite{jenkins,huang,zhang2} values for Bi$_2$Te$_3$. 
We compute the quasi-particle decay rate $\Gamma$ and
find $\Gamma\sim T$ at high temperatures. This prediction
should be observable by ARPES.
Phonons also affect the surface resistivity and yield a characteristic
$T$-dependent contribution to $\rho$.  Our theory is flexible enough 
to allow for the use of microscopically derived phonon modes and 
coupling matrix elements, e.g., resulting from (future) numerical
force-constant calculations.   

\textit{Model.---}
We consider energies below the TI bulk gap where only surface electronic states 
are relevant. For the half-space $z>0$,
the surface-state wave function $\chi(z)$ 
follows from the low-energy band structure with 
Dirichlet boundary conditions at $z=0$ \cite{zhang1},
\begin{equation}\label{chi}
\chi(z) = {\cal N} (e^{-\eta_- z}-e^{-\eta_+ z}),
\quad \eta_\pm = \frac{B_0\pm \sqrt{B_0^2+4M_0M_1}}{2M_1}
\end{equation}
with normalization ${\cal N}$ and 
material parameters $(B_0,M_0,M_1)$ specified 
in Ref.~\cite{zhang2}. Since $M_0 M_1<0$
and $B_0/M_1>0$, we have ${\rm Re}(\eta_\pm) >0$ and the
state (\ref{chi}) decays exponentially.  One arrives at a
massless 2D Dirac Hamiltonian (we set $\hbar=1$) \cite{hasan},
\begin{equation}\label{chiralbasis}
H_e = \sum_{{\bm k},s=\pm} \epsilon_{{\bm k}s} c_{{\bm k}s}^\dagger
c_{{\bm k}s}^{}, \quad \epsilon_{{\bm k}s}= s v_F |{\bm k}| -\mu,
\end{equation}
with the Fermi velocity $v_F\simeq 4.36 \times 10^5$~m/s and the
chemical potential $\mu$ defining $k_F=|\mu|/v_F$.  
A helical eigenstate with helicity
$s=\pm$ has its spin structure tied to the surface momentum 
${\bm k}=(k_x,k_y)$.   Helical fermions, 
$c_{\bm k}=(c_{{\bm k}+},c_{{\bm k}-})^T$,
are connected to the usual spinful operators, $d_{\bm k}=
(d_{{\bm k}\uparrow},d_{{\bm k} \downarrow})^T$, by a 
unitary transformation, 
\begin{equation}\label{unitary}
c_{{\bm k}}  = U_{\bm k}  d_{\bm k}  ,\quad U_{\bm k}=\frac{1}{\sqrt{2}}
\left(\begin{array}{cc} e^{i\theta_{\bm k}/2} & ie^{-i\theta_{\bm k}/2} \\
e^{i\theta_{\bm k}/2} & -ie^{-i\theta_{\bm k}/2} \end{array}\right),
\end{equation}
where $\tan \theta_{\bm k}=k_y/k_x$.

In order to describe noninteracting acoustic phonons
we employ isotropic elastic continuum theory
with stress-free boundary conditions at $z=0$.
We briefly summarize the resulting eigenmodes \cite{landau,sirenko} 
before turning to the electron-phonon coupling. 
Following the notation in Ref.~\cite{sirenko} we label the modes by the 
quantum numbers $\Lambda=({\bm q},\Omega,\lambda)$, 
with surface momentum ${\bm q}=(q_x,q_y)$, 
frequency $\Omega>0$, and mode type $\lambda\in (H,T,L,R)$ explained below.  
In this non-standard but very convenient notation, the
frequency $\Omega=\Omega_\Lambda$ is not specified
in terms of ${\bm q}$ and $\lambda$ but represents a free parameter. 
With ${\bm r}=(x,y)$ and surface area ${\cal A}$,
the displacement field operator takes the form
\begin{equation}\label{udef}
{\bm U}({\bm r},z,t) = \sum_\Lambda 
\frac{1}{\sqrt{2\rho_M {\cal A} \Omega}}
{\bm u}_\Lambda(z) e^{i({\bm q}\cdot {\bm r}-\Omega t)}
 b_\Lambda + {\rm h.c.},
\end{equation}
where $b_\Lambda$ is a bosonic annihilation operator and
$\rho_M\simeq 7860$~kg$/$m$^3$ \cite{jenkins}. 
The noninteracting phonon Hamiltonian 
is $H_{p}=\sum_\Lambda \Omega_\Lambda (b_\Lambda^\dagger b_\Lambda^{}+1/2)$. 
The orthonormal eigenmodes ${\bm u}_\Lambda(z)$  
describe linear combinations of $e^{\pm i k_{l,t} z}$ waves,
where  $k_{l,t} = \sqrt{(\Omega/c_{l,t})^2-q^2}$.
First, the horizontal shear mode, $\lambda=H$,
with ${\bm u}_H  \parallel\hat e_z \times \hat e_q$ (where
$\hat e_q={\bm q}/q$) decouples from all other modes and 
does not generate a deformation potential; hence it is not discussed further.  
The remaining modes are given by
\begin{eqnarray} \nonumber
{\bm u}(z) & =&  \left( iq \phi_l-\frac{d\phi_t}{dz} \right) \hat e_q
+ \left( \frac{d\phi_l}{dz} +iq \phi_t \right)\hat e_z, \\ \label{phitl}
\phi_{l,t} &=& \frac{1}{\sqrt{2\pi \Omega k_{l,t}}} 
\left ( a_{l,t}e^{-ik_{l,t}z} + b_{l,t} e^{ik_{l,t}z} \right).
\end{eqnarray}
The incoming longitudinal mode, $\lambda=L$,
with $a_l=1$ and $a_t=0$, exists for $\Omega>c_l q$ with real
$k_{l,t}>0$. The eigenstate $\phi_{l,t}^{(L)}$ 
has $b_l=-A$ and $b_t=B$, where 
\[
A= \frac{(q^2-k_t^2)^2-4q^2k_lk_t}{(q^2-k_t^2)^2+4q^2k_lk_t},
\quad  B=  \frac{4q(q^2-k_t^2)\sqrt{k_l k_t}}{(q^2-k_t^2)^2+4q^2k_lk_t}.
\]
The incoming transverse mode, $\lambda=T$, with  
$a_l=0$ and $a_t=1$, exists for $\Omega>c_t q$. 
The eigenstate $\phi_{l,t}^{(T)}$ has $b_l=-B$ and $b_t=-A$.
(For $c_t q<\Omega<c_l q$, we have $k_l= i|k_l|$.)
Finally, the energetically lowest solution is the 
Rayleigh surface wave, $\lambda=R$, where $a_l=a_t=0$ 
and $k_{l,t}=i \kappa_{l,t} q$.  Here the dispersion
relation is linear, $\Omega=c_R q$ with surface velocity $c_R= \xi c_t$, 
i.e., $\Omega$ is not a free parameter in $\Lambda$ anymore. 
Putting $\gamma=(c_t/c_l)^2$, we find
\[
\xi = \left( \frac{8}{3}-\frac{4\sqrt{12\gamma-2}}{3}
\cos\left[\frac{1}{3} \cos^{-1}\left(\frac{17-
45\gamma}{(12\gamma-2)^{3/2}}\right)\right] \right)^{1/2}.
\]
With $\kappa_l = \sqrt{1-\gamma \xi^2}$ and $\kappa_t=\sqrt{1-\xi^2}$, 
we obtain
\begin{eqnarray}
\nonumber
\phi^{(R)}_l&=&\sqrt{\frac{C}{q}} \ e^{-\kappa_l qz}, \quad \phi^{(R)}_t=-
\sqrt{\frac{C}{q}} \frac{2i\kappa_l} {1+\kappa_t^2} e^{-\kappa_t qz} , \\
\label{cdef}
C^{-1} &=& \kappa_l-\kappa_t+\frac{(\kappa_l-\kappa_t)^2}{2\kappa_l^2\kappa_t}.
\end{eqnarray}
Using the above values for $c_{l,t}$ we find
$\xi\simeq 0.92$, $\kappa_l\simeq 0.85$, $\kappa_t\simeq 0.39$ and
$C\simeq 1.20$.  

\textit{Electron-phonon coupling.---}
The deformation potential couples the local electron density to 
$\nabla \cdot {\bm U}({\bm r},z)$, 
with a coupling constant $\alpha$. Ref.~\cite{huang} gives
the estimate $\alpha\approx 35$~eV. 
This yields the second-quantized interaction Hamiltonian
\begin{equation}\label{epdef}
H_{ep}= \frac{\alpha}{\sqrt{\cal A}} \sum_{{\bm q}\Omega\lambda} 
M^{(\lambda)}_{q \Omega} \ b^{}_{{\bm q}\Omega\lambda} \sum_{{\bm k}ss'}  
c_{{\bm k}+{\bm q},s}^{\dagger} X_{{\bm k} {\bm q},ss'}
c_{{\bm k}s'}^{} + {\rm h.c.},
\end{equation}
where $U_{\bm k}$ in Eq.~\eqref{unitary} defines the matrix
$X_{\bm k \bm q} = U_{{\bm k}+{\bm q}}^{} U_{\bm k}^\dagger$.
For the Rayleigh mode, the sum over $\Omega$ should
be omitted with the replacement $\Omega=c_R q$.
With $\phi_l^{(\lambda)}$ specified in Eqs.~\eqref{phitl} and \eqref{cdef}, 
we obtain the  $\hat e_q$-independent electron-phonon coupling matrix elements 
\begin{equation}\label{coupl}
M_{q\Omega}^{(\lambda)} = - \frac{(\Omega/c_l)^2}{\sqrt{2\rho_M \Omega}}
\int_0^\infty dz \ |\chi(z)|^2 \phi_l^{(\lambda)}(z),
\end{equation}
with the electronic surface state $\chi$ in Eq.~\eqref{chi}.
For $q\ll {\rm Re}(\eta_-)$, the overlap integral above
reduces to $\phi^{(\lambda)}_l(z=0)$.
In what follows, we discuss physical consequences obtained from
the Hamiltonian $H=H_e+H_p+H_{ep}$.  In the concrete examples below, 
the chemical potential is $\mu=v_F k_F=0.05$~eV,
corresponding to the BG temperature $T_{\rm BG}= 2k_F c_R/k_B=3.9$~K and 
the Fermi temperature $T_F=580$~K.  

\textit{Lifetime broadening.---} We begin with the self-energy
$\Sigma_s({\bm k},\omega)$ for a helical eigenstate $s=\pm$.
Following standard arguments \cite{eiguren,echenique}, the
main contribution is captured to lowest nontrivial order in $H_{ep}$.
Noting that the ``tadpole'' diagram vanishes identically,
the ``rainbow'' diagram \cite{eiguren} yields
independent contributions from each mode $\lambda$,
\begin{eqnarray} \nonumber
\Sigma_{s}^{(\lambda)}({\bm k},\omega) &=& \sum_{s',\nu=\pm} 
\alpha^2 \int_0^\infty d\Omega \int \frac{d^2 {\bm q}}{(2\pi)^2} 
\left|M^{(\lambda)}_{q\Omega} X_{{\bm k}{\bm q},ss'}\right|^2 \\ 
&\times& \label{selfenergy}
 \frac{\nu \left[n_B(\nu \Omega)+n_{F}(\epsilon_{{\bm k}+{\bm q},s'} ) \right]}
{\omega + i0^+ + \nu \Omega- \epsilon_{{\bm k}+{\bm q},s'}},
\end{eqnarray}
where $n_B$ ($n_F$) is the Bose (Fermi) function.
For the $R$ mode, the $q$-integral has to include the additional factor
$\delta(\Omega-c_R q)$, while for $\lambda=L,T$, we have the 
respective constraint $q<\Omega/c_{l,t}$.  The decay rate 
$\Gamma^{(\lambda)}= -2 \ {\rm Im}\ \Sigma^{(\lambda)}$ describing 
lifetime broadening is then given by
\begin{eqnarray} \label{rate}
\Gamma^{(\lambda)}_s({\bm k},\omega)  &=& \sum_{\nu=\pm} \alpha^2 
\int_0^\infty d\Omega \ F^{(\lambda\nu)}_{{\bm k}s,\omega}(\Omega) \\
\nonumber &\times& 
 \left [ n_B(\Omega) +n_F(\Omega+\nu\omega)\right].
\end{eqnarray}
Here the Eliashberg function \cite{eiguren} is defined as
\begin{equation}\label{elias0}
F_{{\bm k}s, \omega}^{(\lambda\pm)}( \Omega ) =\sum_{s'} \int 
\frac{d^2{\bm q}}{2\pi}  
\left|M^{(\lambda)}_{q\Omega} X_{{\bm k}{\bm q},ss'}\right|^2 
 \delta\left(\omega\pm \Omega- \epsilon_{{\bm k}+{\bm q},s'} \right )
\end{equation}
which represents a phonon density of states weighted by the coupling 
matrix elements.
Performing the angular integration yields the result
\begin{eqnarray}\nonumber
F_{ks,\omega}^{(\lambda\nu)}(\Omega ) &=&
\frac{1}{2\pi v_F k}\int_{q_{-}}^{q_{+}}
q dq \left| M^{(\lambda)}_{q\Omega}\right|^2 
\left( \frac{q_{+}^2-q^2}{q^2-q_{-}^2}\right)^{s'/2} ,\\  \label{elias}
q_{\pm} &=& \left|\frac{|\mu+\omega+\nu\Omega|}{v_F} \pm k \right|,
\end{eqnarray}
where $s'\equiv s \ {\rm sgn}(\mu+\omega+\nu\Omega)$.

\begin{figure}[t]
\includegraphics[width=\linewidth,angle=0]{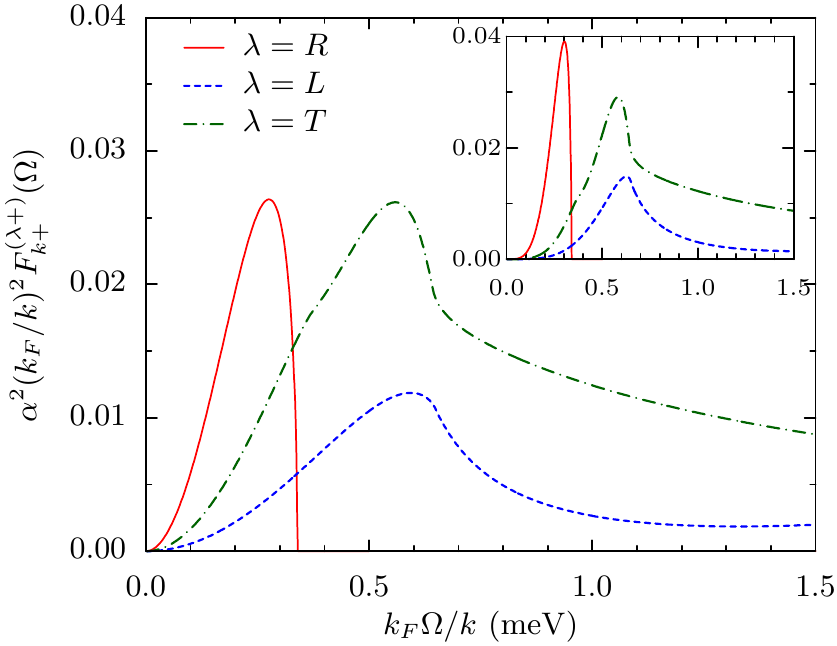}
\caption{\label{fig1} (Color online) Low-frequency
behavior of the Eliashberg functions 
$F^{(\lambda +)}_{k+}(\Omega)$ for the three
relevant acoustic phonon modes ($k=0.1k_F$).  In the rescaled 
units used here, the functions are approximately $k$-independent.
Inset: Same for the ``transport'' Eliashberg function ${\cal F}$ 
(see main text). }
\end{figure}

For a discussion of the lifetime,  we now
consider the on-shell case, $\omega=\epsilon_{{\bm k}s}$. 
For the Rayleigh mode with $c_R\ll v_F$,
we find $s'=+$, $q_{+}\simeq 2k$ and $q_{-}=0$, yielding
for both $\nu=\pm$ the analytical result 
\begin{equation}\label{elias2}
F^{(R)}_{k}(\Omega) = \frac{C}{2\pi} \frac{\Omega^2 
\sqrt{1-(\Omega/2c_R k)^2}}{\rho_M v_F c_l^4} \Theta(2k c_R-\Omega)
\end{equation}
with the Heaviside function $\Theta$. 
The Eliashberg functions for the other two phonon modes have to be
computed numerically.  Together with Eq.~\eqref{elias2} they are shown
in Fig.~\ref{fig1}.  Numerically, after a rescaling we find almost 
\textit{universal}\ behavior
in the sense that the functions $(k_F/k)^2 F_{ks}^{(\lambda\nu)} 
( k\Omega / k_F)$ are essentially independent of $k$. 
The  $\Omega\to 0$ behavior is dominated by the Rayleigh mode 
with $F(\Omega)\sim \Omega^2$,
but at higher energy scales (in particular outside the regime shown in 
Fig.~\ref{fig1}), the two other modes are much more important.

\begin{figure}[t]
\includegraphics[width=\linewidth,angle=0]{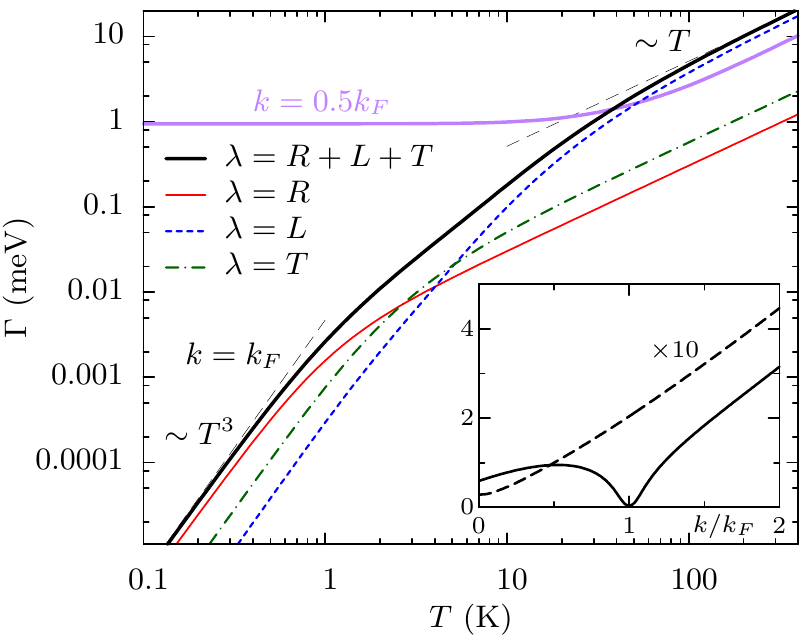}
\caption{\label{fig2} (Color online) Main panel: $T$-dependence of the
 decay rate $\Gamma$ for $k=k_F$ and for $k=0.5 k_F$.  For $k=0.5 k_F$, only the
$L$ mode gives significant contributions.  Inset: $k$-dependence
of $\Gamma$ for $T=3.9$~K (solid line) and for $T=392$~K (dashed line;
the shown result has to be multiplied by 10).  
 }
\end{figure}

The resulting quasi-particle decay rate $\Gamma_k(T)$ 
then follows from Eq.~(\ref{rate}) and is shown in Fig.~\ref{fig2}.   
The decay rate is dominated by the $L$ mode except for very low energy
scales, i.e., when the particle is near the Fermi surface,  $k\approx k_F$,
and temperature is low, $T\alt T_{\rm BG}$.
For high temperatures, however, Eq.~\eqref{rate} predicts
a characteristic $\Gamma\sim T$ law, which allows to 
identify electron-phonon scattering processes in practice.
For $k=k_F$ and $T\ll T_{\rm BG}$, the decay rate is dominated by 
the $R$ mode, and we obtain 
\begin{equation}
\Gamma_{k_F}(T) = \frac{28 \zeta(3) C}{\pi} 
\frac{\alpha^2 c_R^3 k_F^3 }{\rho_M v_F c_l^4} 
\left(\frac{T}{T_{\rm BG}}\right)^3
\end{equation}
with $\zeta(3) \simeq 1.202$ \cite{abramowitz}.
This $T^3$ law and the crossover to the linear $T$ dependence for
$T\gg T_{\rm BG}$ are shown in Fig.~\ref{fig2}. 
We note that away from the Fermi surface, the 
$T=0$ decay rate stays finite and scales as 
$\Gamma_k\sim |k-k_F|^3$ for $k\to k_F$.

\begin{figure}[t]
\includegraphics[width=\linewidth,angle=0]{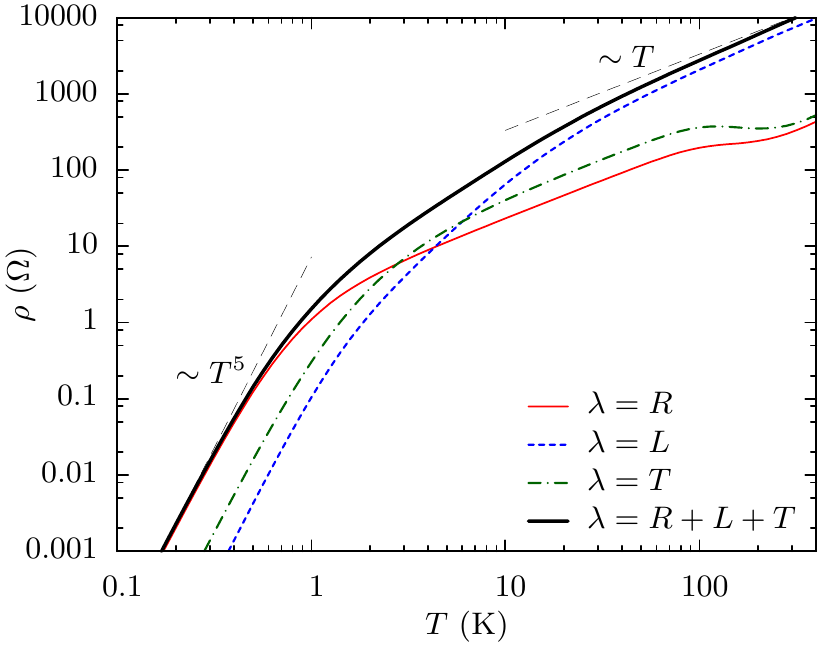}
\caption{\label{fig3} (Color online)  
Temperature dependence of the phonon contribution to the 
surface resistivity.  Note the double-logarithmic scales.
Individual contributions of each phonon mode are also shown. 
}
\end{figure}

\textit{Resistivity.---}
Next we compute the phonon contribution to the resistivity, $\rho$,
using a quasiclassical Boltzmann transport theory as
employed recently for graphene \cite{dassarma, felix},
\begin{equation}\label{resist}
 \rho =  \frac{2}{e^2 v_F^2 D(\mu)} \frac{1}{\langle\tau\rangle},\quad 
\langle \tau \rangle= \frac{\int d\epsilon \ (-\partial_\epsilon n_F) 
 D(\mu+\epsilon)\tau(\epsilon)}{\int d\epsilon\ 
(-\partial_\epsilon n_F) D(\mu+\epsilon)},
\end{equation}
with the density of states $D(E)=|E|/(2\pi v_F^2)$.  
This approach is valid for $|\mu|\langle \tau \rangle \gg 1$, 
which is equivalent to $G_Q\rho\ll 1$ with the conductance 
quantum $G_Q=e^2/h$.
The inverse of the energy-dependent electron-phonon transport scattering 
time $\tau(\epsilon_{{\bm k} s})$ follows from
Fermi's golden rule as a sum over 
independent phonon mode ($\lambda$) contributions.  
The result can again be expressed using a ``transport'' Eliashberg function 
${\cal F}^{(\lambda\pm)}_{{\bm k}s}(\Omega)$ given by Eq.~(\ref{elias0})
with $\omega=\epsilon_{{\bm k}s}$ and
an additional factor $(1-\cos\theta_{{\bm k},{\bm q}})$ in the integral,
where $\theta_{{\bm k},{\bm q}}=\theta_{{\bm k}+{\bm q}}-\theta_{\bm k}$
is the angle between ${\bm k}$ and ${\bm k}+{\bm q}$.  
After the angular integration, we obtain ${\cal F}$ as 
in Eq.~\eqref{elias} but with an additional factor 
$(q^2-q_-^2)/ [2k(q_+-k)]$ in the integrand.
The resulting functions are depicted in the inset of Fig.~\ref{fig1}.
After some algebra, we arrive at \cite{foot1}
\begin{eqnarray}\label{relatim}
\frac{1}{\tau(\epsilon_{{\bm k}s})} & = & \sum_{\lambda, \nu=\pm} 
\alpha^2 \int_0^\infty d\Omega\ 
{\cal F}^{(\lambda,\nu)}_{{\bm k} s}(\Omega)
 \\ \nonumber &\times& \ 
\nu n_B(\nu \Omega) \  \frac{1-n_F(\epsilon_{{\bm k}s}+\nu\Omega)}
{1-n_F(\epsilon_{{\bm k}s})}.
\end{eqnarray}
In Fig.~\ref{fig3}, we show the 
full $T$ dependence of the phonon-induced resistivity $\rho$.
For $T\ll T_{\rm BG}$, the resistivity is dominated
by the $\Omega\to 0$ behavior of ${\cal F}(\Omega)$. 
The latter comes from the $R$ mode with 
${\cal F}\sim \Omega^4$, which
implies  $\rho\sim T^{5}$ as $T\to 0$.  
The prefactor can be evaluated exactly, 
\begin{equation}
G_Q\rho(T\to 0) = \frac{1488 \zeta(5) C}{\pi} 
\frac{\alpha^2 c_R^3 k_F^2}{\rho_M v_F^2 c_l^4} 
\left( \frac{T}{T_{\rm BG}}\right)^5,
\end{equation}
where $\zeta(5)\simeq 1.037$ \cite{abramowitz}.
We  thus recover the standard BG power law, $\rho\sim T^5$, as
in bulk 3D metals \cite{ashcroft} which is
here caused by the coupling to the Rayleigh surface phonon mode.   
For $T\gg T_{\rm BG}$, on the other hand, we find a $\rho\sim T$ law 
predominantly due to the $L$ mode.  For $T\approx T_{\rm BG}$, all 
three phonon modes are important. 

\textit{Conclusions.---} We have formulated an analytically 
tractable effective low-energy theory of the surface Dirac fermion state 
in a strong TI with deformation-potential coupling
to acoustic phonons.  The influence of phonons could be observed as 
characteristic temperature-dependent decay rate $\Gamma$ of 
quasi-particles in ARPES, or from their $T$-dependent contribution to 
the surface resistivity.  The phonon-mediated effective interaction among
surface fermions can also be attractive at low frequencies, possibly 
allowing for superconducting correlations; however, this topic as
well as studies of the electron-induced modification of phonon properties 
in this system or the physics near the Dirac point ($k_F=0$)
are left for future work.  
We hope that our predictions will soon be tested experimentally.

This work was supported by the Humboldt foundation and by
the SFB TR 12 of the DFG.


\begin{thebibliography}{99}

\bibitem{hasan}
M.Z. Hasan and C.L. Kane, Rev. Mod. Phys. {\bf 82}, 3045 (2010).

\bibitem{qizhan}
X.L. Qi and S.C. Zhang, arXiv:1008.2026.

\bibitem{fukane}
L. Fu and C.L. Kane, Phys. Rev. B {\bf 76}, 045302 (2007).

\bibitem{qiz}
X.L. Qi, T.L. Hughes, and S.C. Zhang, Phys. Rev.   B {\bf 78}, 195424 (2008).

\bibitem{arpes}
D. Hsieh \textit{et al.}, Nature {\bf 460}, 1101 (2009);
Y. Chen \textit{et al.}, Science {\bf 329}, 659 (2010);
L.A. Wray \textit{et al.}, Nat. Phys. {\bf 6}, 855 (2010).

\bibitem{butch}
N.P. Butch \textit{et al.}, Phys. Rev. B {\bf 81}, 241301(R) (2010).

\bibitem{ong}
D.X. Qu, Y.S. Hor, J. Xiong, R.J. Cava, and N.P. Ong,
 Science {\bf 329}, 821 (2010).

\bibitem{analytis}
J.G. Analytis \textit{et al.}, Nat. Phys. {\bf 6}, 960 (2010).

\bibitem{richter}
W. Richter, H. K\"ohler, and C.R. Becker, phys. stat. sol. (b) {\bf 84},
619 (1977).

\bibitem{jenkins}
J.O. Jenkins, J.A. Rayne, and R.W. Ure, Jr., Phys. Rev. B {\bf 5}, 3171 (1972).

\bibitem{huang}
B.L. Huang and M. Kaviany, Phys. Rev. B {\bf 77}, 125209 (2008).

\bibitem{shahil}
K.M.F. Shahil, M.Z. Hossain, D. Teweldebrhan, and A.A. Balandin,
Appl. Phys. Lett. {\bf 96}, 153103 (2010); J. Qi \textit{et al.},
\textit{ibid.} {\bf 97}, 182102 (2010).

\bibitem{hofmann}
Ph. Hofmann, Prog. Surf. Sci. {\bf 81}, 191 (2006).

\bibitem{dassarma}
E.H. Hwang and S. Das Sarma, Phys. Rev. B {\bf 77}, 115449 (2008). 

\bibitem{felix}
E. Mariani and F. von Oppen, Phys. Rev. B {\bf 82}, 195403 (2010).

\bibitem{kim}
D.K. Efetov and P. Kim, Phys. Rev. Lett. {\bf 105}, 256805 (2010).

\bibitem{knab}
A. Kn\"abchen, Phys. Rev. B {\bf 55}, 6701 (1997).

\bibitem{eiguren}
B. Hellsing, A. Eiguren, and E.V. Chulkov, J. Phys. Cond. Matt.
{\bf 14}, 5959 (2002).

\bibitem{echenique}
P.M. Echenique \textit{et al.},
Surf. Sci. Rep. {\bf 52}, 219 (2004).  

\bibitem{park}
S.R. Park \textit{et al.}, Phys. Rev. B {\bf 81}, 041405(R) (2010).

\bibitem{landau}
L.D. Landau and E.M. Lifshitz, \textit{Elasticity Theory}, ch.~24
(Pergamon, New York, 1986).

\bibitem{sirenko}
Y.M. Sirenko, K.W. Kim, and M.A. Stroscio, Phys. Rev. B {\bf 56}, 
15770 (1997).

\bibitem{thalmeier}
For a phenomenological discussion of a different electron-phonon coupling
mechanism in the context of surface acoustic waves, see
P. Thalmeier, arXiv:1101.5572.

\bibitem{zhang2}
C.X. Liu \textit{et al.}, 
Phys. Rev. B {\bf 82}, 045122 (2010).

\bibitem{zhang1}
H. Zhang \textit{et al.}, 
Nat. Phys. {\bf 5}, 438 (2009).

\bibitem{abramowitz}
M. Abramowitz and I.A. Stegun, \textit{Handbook of Mathematical
Functions} (Dover, New York, 1971).

\bibitem{foot1}
Equation (\ref{relatim}) ignores screening by the surface charge carriers 
themselves. Note that this approximation is consistent with 
experimental results in graphene \cite{kim}.


\bibitem{ashcroft}
N.W. Ashcroft and N.D. Mermin, \textit{Solid State Physics} (Saunders College,
Philadelphia, 1976).

\end{thebibliography}
\end{document}